\documentclass{aa}
\begin{document}
\newcommand\bsec{\hbox{$.\!\!{\arcsec}$}}
\newcommand\rsec{\hbox{$.\!\!{^{\rm s}}$}}
\newcommand\RA[4]{#1$^{\rm h}$#2$^{\rm m}$#3\rsec#4}
\newcommand\DEC[3]{#1$^{\circ}$#2\arcmin#3\arcsec}
\newcommand{\magpt}[2]{\mbox{$\rm #1\hspace{-0.25em}\stackrel{m}{.}
      \hspace{-1.0mm}#2$}}                             
\newcommand\teff{$T_{\rm eff}$} 
\newcommand\logt{$ \log\,T_{\rm eff}$}
\newcommand\logg{$ \log\,g$}
\newcommand\loghe{$ \log{\frac{n_{\rm He}}{n_{\rm H}}}$}
\newcommand\Ebv{$E_{B-V}$}
\newcommand\ebv{$E_{B-V}$}
\newcommand\logM{$\log {\rm M}$}
\newcommand\Msolar{M$_\odot$}
\thesaurus{05 (08.05.1; 08.06.3; 08.08.2; 10.07.3 NGC~6752)}
\title{Hot HB stars in globular clusters - Physical parameters and 
consequences for theory. V. Radiative levitation versus helium mixing}
\author{S.~Moehler\inst{1}\fnmsep\thanks{Based on observations collected at the 
 European Southern Observatory (ESO~N$^{\b{o}}$~60.E-0145, 61.E-0145,
 61.E-0361).} 
 \and A.V.~Sweigart\inst{2} 
 \and W.B.~Landsman\inst{3}
 \and U.~Heber\inst{1}}
\offprints{S.~Moehler}
\institute {Dr. Remeis-Sternwarte, Astronomisches Institut der Universit\"at
 Erlangen-N\"urnberg, Sternwartstr. 7, 96049 Bamberg, Germany\\
 e-mail: moehler@sternwarte.uni-erlangen.de, heber@sternwarte.uni-erlangen.de
 \and NASA\,Goddard Space Flight Center, Code 681, Greenbelt, 
 MD 20771, USA \\
 e-mail: sweigart@bach.gsfc.nasa.gov,
 \and Raytheon ITSS, NASA\,Goddard Space Flight Center, Code 681, Greenbelt,
 MD 20771, USA \\
 e-mail: landsman@mpb.gsfc.nasa.gov
}
\date{}
\titlerunning{Hot HB stars in globular clusters - V. Radiative 
levitation versus helium mixing }
\maketitle


\begin{abstract}
Atmospheric parameters (\teff, \logg), masses and helium abundances are 
derived for 42 hot horizontal branch (HB) stars in 
the globular cluster NGC\,6752. For
19 stars we derive magnesium and iron abundances as well and find that iron
is enriched by a factor of
50 on average with respect to the cluster abundance
whereas the magnesium abundances are consistent with the cluster abundance.
Radiation pressure may levitate heavy elements like iron to the surface of 
the star in a diffusive process.
Taking into account the enrichment of heavy elements in our spectroscopic
analyses we find that high iron abundances can explain part, but not all, 
of the problem of anomalously low gravities along the blue HB. The 
blue HB stars
cooler than about 15,100~K and the sdB stars (\teff\ $\ge$ 20,000~K) agree
well with canonical theory when analysed with metal-rich ([M/H] = $+$0.5)
model atmospheres, but the stars in between these two groups remain offset
towards lower gravities and masses. Deep Mixing in the red giant
progenitor phase is discussed as another mechanism that may influence the
position of the blue HB stars in the (\teff, \logg)-plane but not their masses.
\end{abstract}


\section{Introduction\label{n6752bhb_sec_intro}} 

The colour-magnitude diagrams of metal-poor globular clusters show a large
variety of horizontal-branch (HB) morphologies, including ``gaps'' along the
blue HB and long ``blue tails'' that extend towards higher effective
temperatures. It has been suggested that the
gaps might separate stars with
different evolutionary origins. Spectroscopic analyses of stars along the blue
HB and blue tails in a number of globular clusters (Moehler \cite{moeh99} and
references therein) yielded the following results: 

\begin{enumerate} \item Most of the stars analysed above and below any gaps are
horizontal branch B  type (HBB) stars ($T_{\rm eff} < 20,000$~K). Their surface
gravities are significantly lower (up to more than 0.5 dex, see Fig.~4 
of Moehler \cite{moeh99}) than expected from canonical HB evolution
theory while their masses are lower than expected by about a factor of 2. For
most clusters the problem of the masses may be solved by new globular cluster
distances derived from {\sc Hipparcos} data (see Reid \cite{reid99} and
references therein, Heber et al.\ \cite{hemo97}, Moehler \cite{moeh99}).

\item Only in NGC~6752 and M~15 have spectroscopic analyses verified the
presence of stars that could be identified with the subdwarf B stars known in
the field of the Milky Way ($T_{\rm eff} > 20,000$~K, \logg\ $>$ 5). In
contrast to the cooler HBB stars these stars show 
gravities and masses that
agree well with the expectations of canonical stellar evolution for extreme HB
stars (Moehler et al.\ \cite{mohe97a}, \cite{mohe97b}).  
\end{enumerate}

There are currently two scenarios for explaining these apparent contradictions:
\begin{description}

\item [{\bf Helium mixing:}]

The dredge-up of nuclearly processed material to the stellar surface of red
giant branch (RGB) stars has been invoked to explain the abundance
anomalies (in C, N, O, Na, and Al) observed in such stars in  many globular
clusters (e.g. Kraft \cite{kraf94}, Norris \& Da Costa \cite{noda95a},
Kraft et al.\ \cite{krsn97}). Since substantial production of Al in
these low-mass stars only seems to occur inside the hydrogen shell (Langer
\& Hoffman \cite{laho95}, Cavallo et al.\ \cite{casw96}, \cite{casw98}), any
mixing process which dredges up Al will also dredge up helium.  Possible
dredge-up mechanisms include rotationally induced mixing (Sweigart \&
Mengel \cite{swme79}, Zahn \cite{zahn92}, Charbonnel \cite{char95}) and
hydrogen shell instabilities (Von Rudloff et al.\ \cite{voru88}, Fujimoto et
al.\ \cite{fuai99}). Such dredge-up would increase the helium abundance in
the red giant's hydrogen envelope and thereby increase the luminosity (and
the mass loss) along the RGB (Sweigart \cite{swei97a}, \cite{swei97b}). 
The progeny of these
stars on the horizontal branch would then have less massive hydrogen
envelopes than unmixed stars. As the temperature of an HB star increases
with decreasing mass of the hydrogen envelope, ``mixed'' HB stars would be
hotter than their canonical counterparts. The helium enrichment would also
lead to an increased hydrogen burning rate and thus to higher luminosities
(compared to canonical HB stars of the same temperature). The luminosities
of stars hotter than about 20,000~K are not affected by this mixing process
because these stars have only inert hydrogen shells. In this
framework the low gravities of hot HB stars would necessarily be connected
to abundance anomalies observed on the RGB, thereby explaining both of
these puzzles at once. 

\item [{\bf Radiative levitation of heavy elements:}] 
Caloi (\cite{calo99}) and Grundahl et al.\ (\cite{grca99}) suggested that
the low surface gravities of the HBB stars are related to a stellar
atmospheres effect caused by the radiative levitation of heavy elements.
Such an enrichment in the metal abundance would change the temperature
structure of the stellar atmosphere and thereby affect the flux
distribution and the line profiles (Leone \& Manfr\`e \cite{lema97}). This
scenario would also account for the fact that there is no evidence for deep
mixing amongst field red giants (e.g. Hanson et al.\ \cite{hasn98},
Carretta et al.\ \cite{cagr99}) even though field HBB stars show the same
low surface gravities as globular cluster stars (Saffer et al.\
\cite{sake97}, Mitchell et al.\ \cite{misa98}). Behr et al.\
(\cite{beco99}, \cite{beco00b}) have recently reported slightly super-solar
iron abundances for HBB stars in M~13 and M~15, in agreement with the
radiative levitation scenario. 

\end{description}

NGC~6752 is an ideal test case for these scenarios: Its distance modulus is
very well determined from both white dwarfs (Renzini et al.\ \cite{rebr96})
and HIPPARCOS parallaxes (Reid \cite{reid97}), and thus any mass
discrepancies cannot be explained by a wrong distance modulus.
Spectroscopic analyses of the faint  blue stars in NGC~6752 showed them to
be subdwarf B (sdB) stars. As mentioned above, their mean mass agrees well
with the canonical value of 0.5~\Msolar. However, almost no stars
in the sparsely populated region above the sdB star region have been
analysed.  If these stars show low surface gravities and canonical masses,
then the combination of deep mixing and the long distance scale (for the
other globular clusters) would resolve the discrepancies described above.
If they show low  surface gravities and low masses, diffusion may indeed
play a r\^ole when analysing these stars for effective temperature and
surface gravity. Then the low surface gravities found for HBB stars could
be artifacts from the use of inappropriate model atmospheres for the
analyses. We therefore decided to observe stars in this region of the
colour-magnitude diagram  and to derive their atmospheric parameters. First
results, which strongly support radiative levitation of heavy elements as
the explanation, have been discussed by Moehler et al.\ (\cite{mosw99a}).
Here we describe the observations and their reductions, provide the
detailed results of the spectroscopic analyses (temperatures, surfaces
gravities, helium and partly iron and magnesium abundances,  masses), and
discuss the consequences of our findings in more detail.

\begin{table*}[h]
\caption[List of target stars in NGC~6752. II]
{Coordinates, photometry, and heliocentric radial velocities for the target
stars in NGC~6752. The numbers refer to Buonanno et al.\ (\cite{buca86}) and
``acc.'' refers to targets that happened to be in the slit by accident. 
We also give the newer $UBV$ photometry of Thompson et al.\ (\cite{thka99}).
For the NTT spectra obtained in 1998 no radial velocities could be determined
due to problems with the wavelength calibration (see Moehler et al.\
\cite{mosw99b}). \label{n6752bhb_targ}}
\begin{tabular}{l|rr|rr|rrr|rc}
\hline
 star & $\alpha_{2000}$ & $\delta_{2000}$ & \multicolumn{2}{c|}{Buonanno et al.\
\cite{buca86} } & \multicolumn{3}{c|}{Thompson et al.\ \cite{thka99}} 
& $v_{\rm rad, hel}$ & acc. \\ 
 &  &  & $V$ & $B-V$ & $V$ & $B-V$ & $U-B$ & [km s$^{-1}$] & \\
\hline
\multicolumn{9}{c}{ESO 1.52m (1998)}\\
\hline
652  & \RA{19}{11}{32}{6} & \DEC{$-$59}{57}{41} & \magpt{14}{70} & 
\magpt{-0}{01}  & \magpt{14}{743} & \magpt{-0}{042} & \magpt{-0}{256} & $-$51
&\\
1132 & \RA{19}{11}{23}{5} & \DEC{$-$59}{58}{14} & \magpt{15}{34} & 
\magpt{-0}{01}  & \magpt{15}{482} & \magpt{-0}{065} &   & $-$53&
\\
1152 & \RA{19}{11}{23}{1} & \DEC{$-$59}{59}{27} & \magpt{15}{21} & 
\magpt{-0}{02}  & \magpt{15}{318} & \magpt{-0}{058} & \magpt{-0}{401} &
$-$54&\\
1157 & \RA{19}{11}{23}{1} & \DEC{$-$59}{56}{36} & \magpt{15}{07} & 
\magpt{+0}{03}  & \magpt{15}{192} & \magpt{-0}{068} & \magpt{-0}{398} &
$-$41&\\
1738 & \RA{19}{11}{09}{7} & \DEC{$-$60}{03}{54} & \magpt{15}{48} & 
\magpt{-0}{03}  & \magpt{15}{561} & \magpt{-0}{090} & \magpt{-0}{458} &
$-$40&\\
2735 & \RA{19}{10}{49}{5} & \DEC{$-$60}{04}{05} & \magpt{14}{43} & 
\magpt{+0}{03}  & \magpt{14}{508} & \magpt{-0}{015} & \magpt{-0}{142} &
$-$66&\\
3253 & \RA{19}{10}{38}{3} & \DEC{$-$59}{51}{38} & \magpt{14}{47} & 
\magpt{+0}{00}  & \magpt{14}{533} & \magpt{-0}{084} & \magpt{-0}{294} &
$-$49&\\  
3348 & \RA{19}{10}{36}{3} & \DEC{$-$60}{00}{15} & \magpt{14}{27} & 
\magpt{+0}{03}  & \magpt{14}{363} &  \magpt{0}{016} &   & $-$27 &
$\times$\\                                        
3408 & \RA{19}{10}{35}{4} & \DEC{$-$60}{00}{19} & \magpt{15}{14} & 
\magpt{+0}{00} & \magpt{15}{238} &  \magpt{0}{062} &   & $-$73 &
 $\times$\\ 
3410 & \RA{19}{10}{35}{3} & \DEC{$-$60}{00}{47} & \magpt{15}{32} & 
\magpt{+0}{02}  & \magpt{15}{397} & \magpt{-0}{093} & \magpt{-0}{471} &
$-$30&\\  
3424 & \RA{19}{10}{35}{1} & \DEC{$-$60}{02}{13} & \magpt{15}{23} & 
\magpt{-0}{05}  & \magpt{15}{360} & \magpt{-0}{036} &   &
$-$51&\\  
3450 & \RA{19}{10}{34}{6} & \DEC{$-$60}{00}{17} & \magpt{14}{81} & 
\magpt{-0}{06}  & \magpt{14}{873} & \magpt{-0}{068} &   &
$-$39&\\  
3461 & \RA{19}{10}{34}{3} & \DEC{$-$60}{01}{50} & \magpt{14}{87} & 
\magpt{-0}{03}  & \magpt{14}{996} & \magpt{-0}{068} &   &
$-$41&\\  
3655 & \RA{19}{10}{30}{2} & \DEC{$-$59}{57}{27} & \magpt{16}{40} & 
\magpt{-0}{22}  & \magpt{16}{425} & \magpt{-0}{168} &   & $-$29 &
$\times$\\
3736 & \RA{19}{10}{28}{3} & \DEC{$-$60}{00}{48} & \magpt{14}{47} & 
\magpt{+0}{07}  & \magpt{14}{598} &  \magpt{0}{003} &   & $-$57 &
$\times$\\
4172 & \RA{19}{10}{19}{1} & \DEC{$-$59}{57}{26} & \magpt{14}{48} & 
\magpt{-0}{04}  & \magpt{14}{536} & \magpt{-0}{063} &   &
$-$50&\\  
4424 & \RA{19}{10}{14}{2} & \DEC{$-$59}{55}{23} & \magpt{14}{70} & 
\magpt{+0}{00}  & \magpt{14}{790} & \magpt{-0}{066} & \magpt{-0}{251} &
$-$55&\\  
4551 & \RA{19}{10}{10}{9} & \DEC{$-$60}{03}{50} & \magpt{14}{93} & 
\magpt{+0}{01}  & \magpt{14}{979} & \magpt{-0}{090} & \magpt{-0}{437} &
$-$55&\\
4822 & \RA{19}{10}{01}{9} & \DEC{$-$60}{01}{12} & \magpt{15}{06} & 
\magpt{+0}{01}  & \magpt{15}{205} & \magpt{-0}{075} & \magpt{-0}{392} &
$-$30&\\  
4951 & \RA{19}{09}{55}{0} & \DEC{$-$60}{01}{25} & \magpt{15}{38} & 
\magpt{-0}{05}  & \magpt{15}{411} & \magpt{-0}{074} & \magpt{-0}{423} &
$-$36&\\  
\hline
 \multicolumn{9}{c}{ESO NTT 1997 (see Moehler et al.\ \cite{mola99})}\\
\hline
944  & \RA{19}{11}{26}{7} & \DEC{$-$59}{56}{03} & \magpt{14}{58} & 
\magpt{+0}{02}  & \magpt{14}{574} & \magpt{-0}{015} & \magpt{-0}{089} & $+$72 &
$\times$\\
1391 & \RA{19}{11}{11}{8} & \DEC{$-$59}{55}{35} & \magpt{15}{84} & 
\magpt{-0}{11}  & \magpt{15}{887} & \magpt{-0}{083} &   & $-$32 &
\\
1780 & \RA{19}{11}{09}{0} & \DEC{$-$59}{52}{06} & \magpt{15}{77} & 
\magpt{-0}{07}  & \magpt{15}{854} & \magpt{-0}{134} & \magpt{-0}{483} & $-$32 &
\\
2099 & \RA{19}{11}{03}{3} & \DEC{$-$59}{55}{39} & \magpt{15}{88} & 
\magpt{-0}{09}  & \magpt{15}{909} & \magpt{-0}{091} &   & $-$38 &
\\
\hline
\multicolumn{9}{c}{ESO NTT 1998 (see Moehler et al.\ \cite{mosw99b})}\\
\hline
2697 & \RA{19}{10}{50}{3} & \DEC{$-$60}{01}{21} & \magpt{14}{29} & 
\magpt{+0}{08}  & \magpt{14}{291} &  \magpt{0}{025} & \magpt{-0}{043} & -- &
$\times$\\
2698 & \RA{19}{10}{50}{3} & \DEC{$-$60}{02}{34} & \magpt{15}{08} & 
\magpt{-0}{13}  & \magpt{15}{202} & \magpt{-0}{056} &   & -- & \\   
2747 & \RA{19}{10}{49}{1} & \DEC{$-$59}{52}{55} & \magpt{15}{99} & 
\magpt{-0}{07}  & \magpt{16}{061} & \magpt{-0}{078} &   & -- & \\   
2932 & \RA{19}{10}{44}{9} & \DEC{$-$59}{51}{48} & \magpt{16}{08} & 
\magpt{-0}{11}  & \magpt{16}{105} & \magpt{-0}{082} &   & -- & \\   
3006 & \RA{19}{10}{43}{4} & \DEC{$-$59}{56}{56} & \magpt{16}{14} & 
\magpt{-0}{10}  & & & & -- & \\
3094 & \RA{19}{10}{41}{5} & \DEC{$-$60}{02}{51} & \magpt{14}{19} &
\magpt{+0}{03}  & \magpt{14}{222} &  \magpt{0}{028} &   & -- &
$\times$\\
3140 & \RA{19}{10}{40}{7} & \DEC{$-$59}{51}{55} & \magpt{13}{97} & 
\magpt{+0}{08}  & \magpt{13}{975} &  \magpt{0}{058} &  \magpt{0}{116} & -- &
$\times$\\
3253 & \RA{19}{10}{38}{3} & \DEC{$-$59}{51}{38} & \magpt{14}{47} & 
\magpt{+0}{00}  & \magpt{14}{533} & \magpt{-0}{084} & \magpt{-0}{294} & -- & \\   
3699 & \RA{19}{10}{29}{3} & \DEC{$-$59}{58}{11} & \magpt{16}{05} & 
\magpt{-0}{07}  & \magpt{16}{115} & \magpt{-0}{123} &   & -- & \\   
\hline
\multicolumn{9}{c}{ESO NTT 1993 (see Moehler et al.\ \cite{mohe97b})}\\
\hline
491 & \RA{19}{11}{37}{7} & \DEC{$-$60}{03}{11} & \magpt{17}{45} & \magpt{-0}{31} 
 & \magpt{17}{387} & \magpt{-0}{231} & \magpt{-0}{805} & $+$7 & \\
916 & \RA{19}{11}{27}{3} & \DEC{$-$60}{03}{51} & \magpt{17}{61} & \magpt{-0}{26} 
 & \magpt{17}{544} & \magpt{-0}{188} & \magpt{-0}{792} & $+$12 & \\
1509 &\RA{19}{11}{15}{0} & \DEC{$-$59}{54}{31} & \magpt{15}{52} & \magpt{-0}{06} 
 & \magpt{15}{535} & \magpt{-0}{086} & & $-$33 & \\
1628 &\RA{19}{11}{11}{7} & \DEC{$-$59}{59}{36} & \magpt{16}{30} & \magpt{-0}{17} 
 & \magpt{16}{249} & \magpt{-0}{064} & & $-$21 & \\
2162 & \RA{19}{11}{01}{9} & \DEC{$-$60}{03}{02} & \magpt{17}{88} & \magpt{-0}{27} 
 & \magpt{17}{797} & \magpt{-0}{210} & \magpt{-0}{885} & $-$36 & \\
2395 & \RA{19}{10}{57}{3} & \DEC{$-$60}{03}{32} & \magpt{16}{73} & \magpt{-0}{24} 
 & \magpt{16}{715} & \magpt{-0}{147} & & $-$52 & \\
3915 & \RA{19}{10}{24}{2} & \DEC{$-$59}{54}{22} & \magpt{17}{16} & \magpt{-0}{23} 
 & \magpt{17}{183} & \magpt{-0}{215} & \magpt{-0}{823} & $-$9 & \\
3975 &\RA{19}{10}{22}{7} & \DEC{$-$60}{03}{01} & \magpt{16}{27} & \magpt{-0}{22} 
 & \magpt{16}{310} & \magpt{-0}{135} & \magpt{-0}{602} & $+$18 & \\
4009 & \RA{19}{10}{22}{2} & \DEC{$-$60}{07}{47} & \magpt{17}{44} & \magpt{-0}{30} 
 & \magpt{17}{395} & \magpt{-0}{228} & \magpt{-0}{853} & $-$56 & \\
4548 & \RA{19}{10}{10}{9} & \DEC{$-$59}{51}{31} & \magpt{16}{60} & \magpt{-0}{14} 
 & \magpt{16}{653} & \magpt{-0}{161} & \magpt{-0}{616} & $-$65 & \\
\hline
\end{tabular}
\end{table*}

\begin{figure}
\vspace{12.cm}
\includegraphics{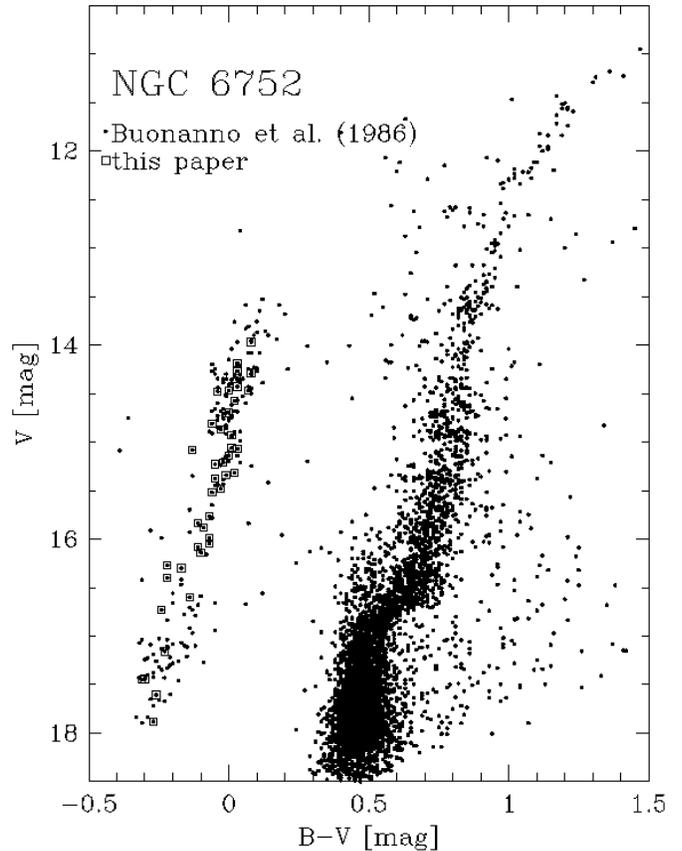}
\caption[CMD of NGC~6752. II]
{The colour-magnitude diagram of NGC 6752 (Buonanno et al.\ \cite{buca86}).
Stars analysed in this paper (including 10 stars discussed by Moehler et
al.\ \cite{mohe97b}) are marked by open squares (some of which overlap due
to almost identical photometric data). \label{n6752bhb_cmd}} 
\end{figure}

\section{Observations\label{n6752bhb_sec_obs}} We selected our targets from
the photographic photometry of Buonanno et al.\ (\cite{buca86}, see
Table~\ref{n6752bhb_targ} and Fig.~\ref{n6752bhb_cmd}). For our
observations we used the ESO 1.52m telescope with the Boller \& Chivens
spectrograph and CCD \#39 (2048$\times$2048 pixels, (15~$\mu$m)$^2$ pixel
size, read-out noise 5.4~e$^-$, conversion factor 1.2~e$^-$/count). We used
grating \# 33 (65~\AA mm$^{-1}$) to cover a wavelength of 3300~\AA\ -- 5300~\AA .
Combined with a slit width of 2\arcsec\ we thus achieved a spectral
resolution of 2.6~\AA . The spectra were obtained on July 22-25, 1998. For
calibration purposes we observed each night ten bias frames and ten dome
flat-fields with a mean exposure level of about 10,000~counts 
each. Before and
after each science observation we took HeAr spectra for wavelength
calibration purposes. We observed dark frames of 3600 and 1800 sec duration
to measure the dark current of the CCD. As flux standard stars we used
LTT~7987 and EG~274. 

We also analyse data that were obtained as backup targets at the NTT during
observing runs dedicated to other programs (60.E-0145, 61.E-0361). The
observational set-up and the data reduction are described in Moehler et al.\
(\cite{mola99}, \cite{mosw99b}). These data have a much lower resolution of
5.4~\AA . 

\section{Data Reduction\label{n6752bhb_sec_redu}}

\begin{figure*}
\vspace{22cm}
\includegraphics{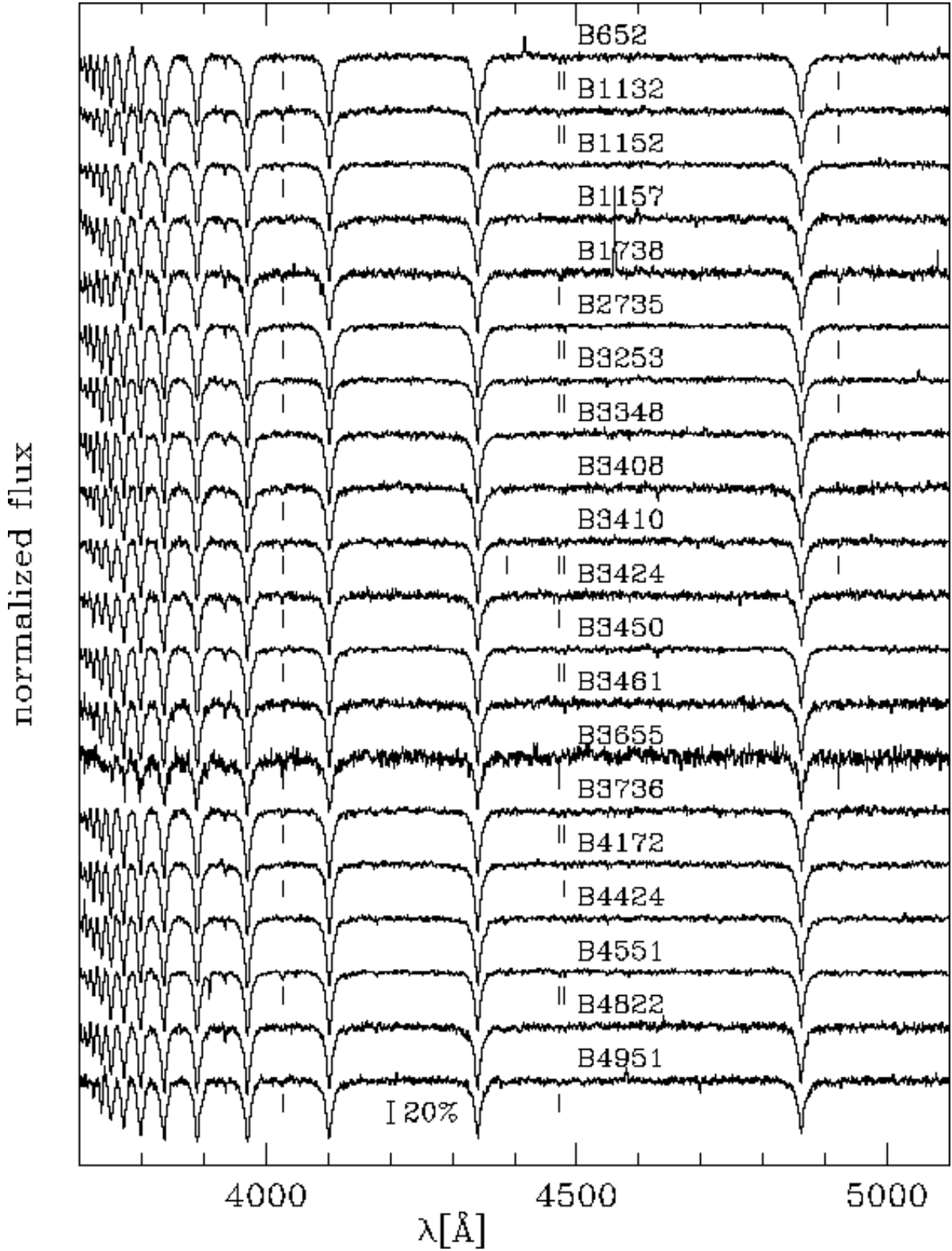}
\caption{Normalized spectra of the programme stars that were observed at 
the ESO 1.52m telescope. The part shortward of 3900~\AA\ was normalized by 
taking the highest flux point as continuum value. The \ion{He}{i} lines
$\lambda\lambda$ 4026~\AA, 4388~\AA, 4471~\AA, 4922~\AA, and the
\ion{Mg}{ii} line 4481~\AA\ are marked (if visible in the
spectrum).\label{n6752bhb_spec1p5}}
\end{figure*}

\begin{figure*}
\vspace{14.4cm}
\includegraphics{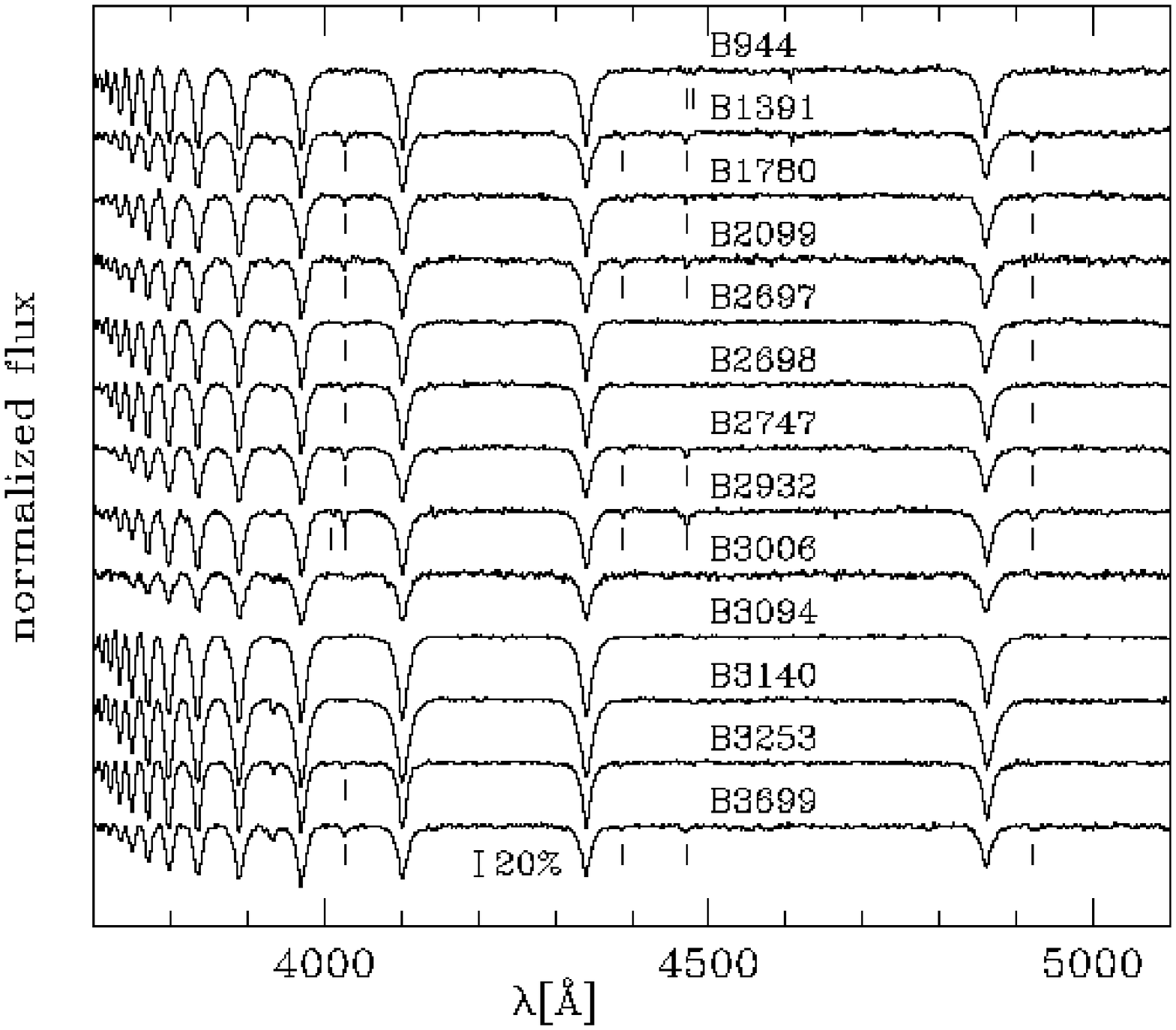}
\caption{Normalized spectra of the programme stars that were observed at 
the NTT during 1997 and 1998. See Fig.~\ref{n6752bhb_spec1p5} for details.
\label{n6752bhb_specntt}}
\end{figure*}

We first averaged the bias and flat field frames separately for each night.
As we could not detect any significant change in the mean bias level we
computed the median of the bias frames of the four nights and found that
the bias level showed a gradient across the image, increasing from the
lower left corner to the upper right corner by about 1\%. We fitted the
bias with a linear approximation along both axes and used this fit as a bias
for the further reduction. As no overscan was recorded we could not adjust
the bias level. Bias frames taken during the night, however, revealed no
significant change in the mean bias level. The mean dark current determined
from long dark frames showed no structure and turned out to be negligible
(3$\pm$3 e$^-$/hr/pixel). 

We determined the spectral energy distribution of the flat field lamp by
averaging the mean flat fields of each night along the spatial axis. These
one-dimensional ``flat field spectra" were then heavily smoothed and used
afterwards to normalize the dome flats along the dispersion axis. The
normalized flat fields of the first three nights were combined. For the
fourth night we used only the flat field obtained during that night as we
detected a slight variation in the fringe patterns of the flat fields from
the first three nights compared to that of the fourth (below 5\%). 

For the wavelength calibration we fitted 3$^{\rm rd}$-order polynomials to
the dispersion relations of the HeAr spectra which resulted in mean
residuals of $\le$0.1~\AA. We rebinned the frames two-dimensionally to
constant wavelength steps. Before the sky fit the frames were smoothed
along the spatial axis to erase cosmic ray hits in the background. To
determine the sky background we had to find regions without any stellar
spectra, which were sometimes not close to the place of the object's
spectrum. Nevertheless the flat field correction and wavelength calibration
turned out to be good enough that a linear fit to the spatial distribution
of the sky light allowed the sky background at the object's position to be
reproduced with sufficient accuracy. This means in our case that after the
fitted sky background was subtracted from the unsmoothed frame we do not
see any absorption lines caused by the predominantly red stars of the
clusters. The sky-subtracted spectra were extracted using Horne's
(\cite{horn86}) algorithm as implemented in MIDAS (Munich Image Data
Analysis System). 

Finally the spectra were corrected for atmospheric extinction using the
extinction coefficients for La Silla (T\"ug \cite{tueg77}) as 
implemented in MIDAS.
The data for the flux standard stars were
taken from Hamuy et al.\ (\cite{hamu92}) and the response curves were fitted
by splines. The flux-calibration is helpful for the later 
normalization of the spectra as it takes out all large-scale sensitivity 
variations of the instrumental setup. Absolute photometric accuracy is not 
an issue here.

\begin{table}[!ht]
\caption[Physical parameters, helium abundances, and masses for targets in
NGC~6752 using metal-poor model atmospheres]
{Physical parameters, helium abundances, and masses for the target stars in
NGC~6752 as derived using metal-poor model atmospheres. We used the
photometry of Buonanno et al.\ (\cite{buca86}) to derive the
masses.\label{n6752bhb_parm15}} 
\begin{tabular}{l|lrrr}
\hline
Star & \multicolumn{1}{c}{\teff} & \multicolumn{1}{c}{\logg} &
 \multicolumn{1}{c}{\loghe} & \multicolumn{1}{c}{M}\\
 & \multicolumn{1}{c}{[K]} & \multicolumn{1}{c}{[cm s$^{-2}$]} & &
\multicolumn{1}{c}{[\Msolar]} \\
\hline
\multicolumn{5}{c}{ESO 1.52m telescope observations in 1998}\\
\hline
  652 & 12500$\pm$310 &  3.86$\pm$0.09 & $-$2.00$\pm$0.35 & 0.55\\
 1132 & 17300$\pm$520 &  4.31$\pm$0.09 & $-$2.46$\pm$0.16 & 0.50\\
 1152 & 15700$\pm$360 &  4.19$\pm$0.05 & $-$2.57$\pm$0.17 & 0.50\\
 1157 & 15800$\pm$460 &  4.14$\pm$0.09 & $-$2.89$\pm$0.31 & 0.50\\
 1738 & 16700$\pm$700 &  4.15$\pm$0.12 & $-$2.24$\pm$0.28 & 0.32\\
 2735 & 11100$\pm$260 &  3.78$\pm$0.12 & $-$1.14$\pm$0.36 & 0.73\\
 3253 & 13700$\pm$390 &  3.80$\pm$0.09 & $-$2.41$\pm$0.29 & 0.50\\
 3348 & 12000$\pm$270 &  3.73$\pm$0.07 & $-$2.18$\pm$0.38 & 0.65\\
 3408 & 14600$\pm$400 &  4.21$\pm$0.09 & $-$2.40$\pm$0.36 & 0.63\\
 3410 & 15500$\pm$460 &  4.14$\pm$0.09 & $-$2.22$\pm$0.19 & 0.41\\
 3424 & 17900$\pm$570 &  4.23$\pm$0.09 & $-$2.60$\pm$0.21 & 0.44\\
 3450 & 13200$\pm$290 &  3.84$\pm$0.07 & $-$2.05$\pm$0.24 & 0.43\\
 3461 & 15200$\pm$500 &  4.18$\pm$0.09 & $\le-$3 & 0.70\\
 3655 & 25800$\pm$1300 &  5.15$\pm$0.16 & $-$2.32$\pm$0.24 & 0.68\\
 3736 & 13400$\pm$370 &  3.91$\pm$0.09 & $-$1.84$\pm$0.17 & 0.67\\
 4172 & 12200$\pm$260 &  3.68$\pm$0.07 & $-$2.24$\pm$0.54 & 0.46\\
 4424 & 13000$\pm$290 &  3.99$\pm$0.07 & $-$2.36$\pm$0.38 & 0.69\\
 4551 & 15400$\pm$530 &  3.96$\pm$0.09 & $-$2.21$\pm$0.24 & 0.39\\
 4822 & 13900$\pm$450 &  3.91$\pm$0.09 & $-$2.24$\pm$0.28 & 0.37\\
 4951 & 17300$\pm$580 &  4.38$\pm$0.09 & $-$2.63$\pm$0.22 & 0.56\\
\hline
\multicolumn{5}{c}{ESO NTT observations in 1997} \\
\hline
  944 & 11100$\pm$230 &  3.70$\pm$0.10 & $-$0.84$\pm$0.31 & 0.52\\
 1391 & 19700$\pm$570 &  4.49$\pm$0.09 & $-$2.04$\pm$0.10 & 0.39\\
 1780 & 18000$\pm$580 &  4.40$\pm$0.09 & $-$2.31$\pm$0.14 & 0.39\\
 2099 & 20000$\pm$820 &  4.61$\pm$0.12 & $-$2.38$\pm$0.22 & 0.48\\
\hline
\multicolumn{5}{c}{ESO NTT observations in 1998}\\
\hline
 2697 & 15700$\pm$400 &  4.08$\pm$0.07 & $-$2.36$\pm$0.17 & 0.91\\
 2698 & 15400$\pm$610 &  4.11$\pm$0.10 & $-$2.07$\pm$0.28 & 0.49\\
 2747 & 22700$\pm$650 &  4.85$\pm$0.09 & $-$2.16$\pm$0.10 & 0.61\\
 2932 & 18600$\pm$700 &  4.63$\pm$0.12 & $-$1.57$\pm$0.12 & 0.47\\
 3006 & 30000$\pm$640 &  5.19$\pm$0.09 & $\le-$3 & 0.71\\
 3094 & 10400$\pm$120 &  3.81$\pm$0.17 & $-$1.83$\pm$1.35 & 1.09\\
 3140$^1$ &  8000$\pm$100 &  2.84$\pm$0.14 & $-$1.00$\pm$0.00 & 0.28\\
 3253 & 13700$\pm$470 &  3.75$\pm$0.10 & $-$1.85$\pm$0.31 & 0.45\\
 3699 & 22900$\pm$990 &  4.64$\pm$0.12 & $-$2.29$\pm$0.10 & 0.35\\
\hline
\multicolumn{5}{c}{ESO NTT observations in 1993}\\
\hline
  491 & 29000$\pm$520 &  5.41$\pm$0.07 & $\le-$3 & 0.38\\
  916 & 30200$\pm$430 &  5.61$\pm$0.07 & $-$1.71$\pm$0.05 & 0.48\\
 1509 & 17400$\pm$630 &  4.10$\pm$0.10 & $-$2.17$\pm$0.16 & 0.26\\
 1628 & 21800$\pm$590 &  4.83$\pm$0.09 & $-$2.53$\pm$0.12 & 0.47\\
 2162 & 33400$\pm$390 &  5.78$\pm$0.07 & $-$1.94$\pm$0.09 & 0.45\\
 2395 & 22200$\pm$690 &  5.10$\pm$0.09 & $-$1.78$\pm$0.07 & 0.57\\
 3915 & 31300$\pm$510 &  5.55$\pm$0.09 & $\le-$3 & 0.59\\
 3975 & 21700$\pm$460 &  4.97$\pm$0.07 & $-$2.04$\pm$0.10 & 0.67\\
 4009 & 30700$\pm$920 &  5.61$\pm$0.12 & $\le-$3 & 0.54\\
 4548 & 22000$\pm$1380 &  5.11$\pm$0.19 & $-$2.02$\pm$0.16 & 0.67\\
\hline
\end{tabular}

\begin{tabular}{ll}
$^1$&This star is omitted from further analysis as it lies in\\
 & a temperature range that is difficult to analyse and  \\
 & not of great interest for our discussion.\\
\end{tabular}
\end{table}
\section{Atmospheric Parameters\label{n6752bhb_sec_par}}

To derive effective temperatures, surface gravities and helium abundances
we fitted the observed Balmer and helium lines with stellar model
atmospheres. Beforehand we corrected the spectra for radial velocity
shifts, derived from the positions of the Balmer and helium lines. The
resulting heliocentric velocities are listed in Table~\ref{n6752bhb_targ}.
The error of the velocities (as estimated from the scatter of the
velocities derived from individual lines) is about 40~km~s$^{-1}$. The spectra
were then normalized by eye and are plotted in Figs.~\ref{n6752bhb_spec1p5}
and \ref{n6752bhb_specntt}. 

To establish the best fit we used the routines developed by Bergeron et al.\
(\cite{besa92}) and Saffer et al.\ (\cite{sabe94}), which employ a $\chi^2$
test. The $\sigma$ necessary for the calculation of $\chi^2$ is estimated
from the noise in the continuum regions of the spectra. The fit program
normalizes model spectra {\em and} observed spectra using the same points
for the continuum definition. 

We computed model atmospheres using ATLAS9 (Kurucz \cite{kuru91}) and used
Lemke's version\footnote{For a description see
http://a400.sternwarte.uni-erlangen.de/$\sim$ai26/linfit/linfor.html} of
the LINFOR program (developed originally by Holweger, Steffen, and
Steenbock at Kiel University) to compute a grid of theoretical spectra
which include the Balmer lines H$_\alpha$ to H$_{22}$ and \ion{He}{i}
lines. The grid covered the range 7,000~K~$\leq$~\teff~$\leq$~35,000~K,
2.5~$\leq$~\logg~$\leq$~6.0, $-3.0$~$\leq$~\loghe~$\leq$~$-1.0$, at a
metallicity of [M/H]~=~$-1.5$. In Table~\ref{n6752bhb_parm15} we list the
results obtained from fitting the Balmer lines H$_\beta$ to H$_{10}$
(excluding H$_\epsilon$ to avoid the \ion{Ca}{ii}~H line) and the
\ion{He}{i} lines 4026~\AA, 4388~\AA, 4471~\AA, and 4921~\AA. The errors
given are r.m.s. errors derived from the $\chi^2$ fit (see Moehler et al.\
\cite{mosw99b} for more details). These errors are obtained under the
assumption that the only error source is statistical noise (derived from
the continuum of the spectrum). However, errors in the normalization of the
spectrum, imperfections of flat field/sky background correction, variations
in the resolution (e.g. due to seeing variations when using a rather large
slit width) and other effects may produce systematic rather than statistic
errors, which are not well represented by the error obtained from the fit
routine. Systematic errors can only be quantified by comparing truly
independent analyses of the same stars. As this is not possible here we use
our experience with the analysis of similar stars and estimate the true
errors to be about 10\% in \teff\ and 0.15~dex in \logg\ (cf. Moehler et
al.\ \cite{mohe97b}, \cite{mola98}). Two stars show $B-V$ colours that are
significantly redder than expected from their effective temperatures
(B~2697: $B-V$ = \magpt{+0}{08}, \teff\ = 15,700~K; B~3006: $B-V$ =
\magpt{-0}{10}, \teff\ = 30,000~K), possibly indicating that the colours
are affected by binarity or photometric blending with a cool star. While
the spectra look quite normal, we will not include these stars in any
statistical discussion below. To increase our data sample we reanalysed the
NTT spectra described and analysed by Moehler et al.\ (\cite{mohe97b}). We
did not reanalyse the EFOSC1 data published in the same paper as they are
of worse quality. We find that the atmospheric parameters determined by
line profile fitting agree rather well with those published by Moehler et
al.\ (\cite{mohe97b}). 

\begin{figure}
\vspace{11.5cm}
\includegraphics{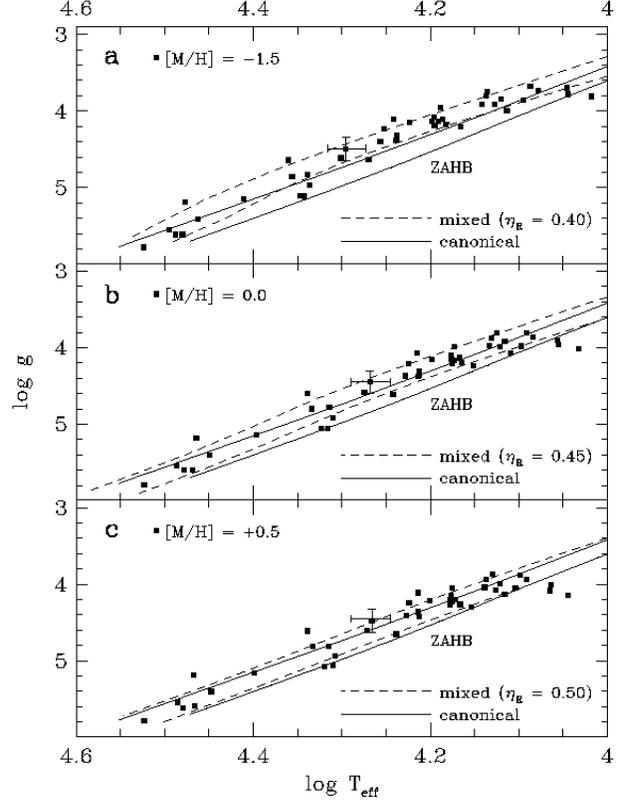}
\caption{{\bf a-c}. 
Temperatures and gravities of the programme stars in NGC~6752.
{\bf a} determined using model atmospheres with cluster
metallicity ([M/H] = $-$1.5), {\bf b} adopting a solar
metallicity ([M/H] = 0) for the model atmospheres, {\bf c}
adopting a super-solar metallicity ([M/H] = $+$0.5) for the model
atmospheres (see Sect.~\ref{n6752bhb_ssec_iron} for details). The
dashed lines mark the locus of the HB evolutionary tracks
for [M/H] = $-$1.54, as computed with helium mixing for
the indicated values of the Reimers mass-loss parameter
$\eta_R$ (see Sect.~\ref{ssec-hemix} for details).  The solid lines
mark the locus of canonical HB tracks for [M/H] = $-$1.54.  These
loci define the region within which the HB models spend 99 percent
of their HB lifetime. Representative error bars are
plotted. \label{n6752bhb_irontg}} 
\end{figure}

The temperatures and
gravities obtained from these metal-poor atmospheres are compared
with the values predicted by canonical HB tracks in
Fig.~\ref{n6752bhb_irontg} (top panel). These tracks,
which were computed for a main sequence mass of 0.805~\Msolar,
an initial helium abundance Y
of 0.23 and a scaled-solar metallicity [M/H]
of $-$1.54, define the locus of canonical HB models
which lose varying amounts of mass during the RGB
phase. According to the Reimers mass-loss formulation the value
of the mass-loss
parameter $\eta_R$ would vary from $\approx$0.4 at the red end
of the observed HB in NGC~6752 to $\approx$0.7
for the sdB stars, given the present composition parameters.

One can see from Fig.~\ref{n6752bhb_irontg} (top panel) that the HBB stars in
NGC~6752 show the same effect as seen in other globular clusters, namely,
an offset from the zero-age horizontal branch (ZAHB) towards lower surface
gravities over the temperature range 4.05 $<$ \logt\ $<$ 4.30 (11,200~K $<$
\teff\ $<$ 20,000~K). At lower or higher temperatures the gravities agree
with the locus of the canonical HB tracks.

\subsection{Radiative levitation of heavy elements\label{n6752bhb_ssec_iron}}

\begin{figure}
\vspace{11.5cm}
\includegraphics{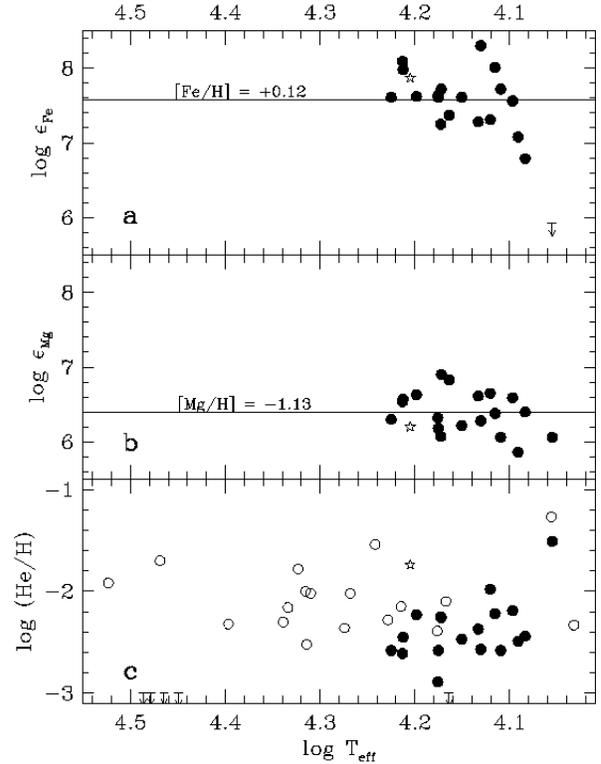}
\caption{{\bf a-c} Abundances of iron ({\bf a}), magnesium ({\bf b}), and
helium ({\bf c}) for the programme stars in NGC~6752. The filled
symbols mark stars which have been observed at higher resolution at the ESO
1.52m telescope, the open symbols mark stars observed at the NTT. Only the
helium abundance could be derived for the NTT stars due to the low
resolution of the data. The asterisk marks the results of Glaspey et al.\
(\cite{glmi89}) for an HBB star in NGC~6752. Upper limits are marked 
by arrows.\label{n6752bhb_abu}} 
\end{figure}

As described in Moehler et al.\ (\cite{mosw99a}, see also 
Fig.~\ref{n6752bhb_abu}), we found evidence
for iron enrichment in the spectra of the HBB stars obtained at the ESO
1.52m telescope, whereas the magnesium abundance appeared consistent with
the cluster magnesium abundance. The actual iron abundances derived for
these stars by fitting the iron lines in the ESO 1.52m spectra are listed in
Table~\ref{irontab}. The mean iron abundance turns out to be [Fe/H] =
$+0.12\pm0.40$ (internal errors only, 
$\log \epsilon_{Fe} = 7.58$) for stars hotter than about
11,500~K -- in good agreement with the findings of Behr et al.\
(\cite{beco99}, \cite{beco00b}) for HBB/HBA (horizontal branch A type)
stars in M~13 and M~15 and Glaspey et al.\ (\cite{glmi89}) for two HBB/HBA stars
in NGC~6752. This iron abundance is a factor of
50 greater than that of the cluster, but still a factor of 3 smaller
than that required to explain the Str\"omgren $u$-jump
discussed by Grundahl et al.\ (\cite{grca99}, $\log \epsilon_{Fe} = 8.1$).
The mean magnesium abundance for the same stars is [Mg/H] = 
$-1.13\pm0.29$ (internal errors only), 
corresponding to [Mg/Fe] = $+$0.4 for [Fe/H] = $-$1.54.  This value
agrees well with the abundance [Mg/Fe] = $+0.4$ found by Norris \& da Costa
(\cite{noda95b}) for red giants in NGC~6752. 

The abundances are plotted versus temperature in Fig.~\ref{n6752bhb_abu}.
The trend of decreasing helium abundance with increasing temperature seen
in the ESO 1.52m data (and also reported by Behr et al.\ \cite{beco99} for HB
stars in M~13) is not supported towards higher temperatures by the NTT
data. This could be due to the lower resolution of the NTT data which may
tend to overestimate abundances (Glaspey et al.\ \cite{glmi89}). 

\begin{figure}
\vspace{6.5cm}
\includegraphics{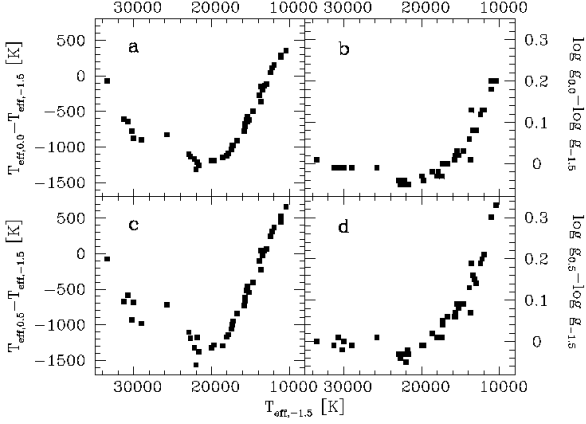}
\caption{{\bf a-d} This plot shows the differences in effective temperature 
({\bf a,c}) and surface gravity 
({\bf b,d}) derived from fits with model atmospheres
of different metallicity (solar$-$metal-poor [{\bf a,b}],
metal-rich$-$metal-poor [{\bf c,d}]). It is obvious that an increase in
the metallicity of the model atmospheres usually decreases the resulting
temperatures and increases the resulting surface gravities.
\label{difference}} 
\end{figure}

As iron is very important for the temperature stratification of stellar
atmospheres we tried to take the increased iron abundance into account by
computing model atmospheres for [M/H] = 0. Indeed a backwarming effect of
2--4\% on the temperature structure was found in the formation region of
the Balmer lines, when comparing solar composition models with the
metal-poor models. We then repeated the fit to derive \teff , \logg , and
\loghe\ with these enriched model atmospheres. The resulting effective
temperatures and gravities changed as displayed in Fig.~\ref{difference}.
The results are listed in Table~\ref{n6752bhb_parp00} and plotted in
Fig.~\ref{n6752bhb_irontg} (central panel).  From
Fig.~\ref{n6752bhb_irontg} (central panel) it is clear that the use of
solar-metallicity model atmospheres moves most stars closer to the
canonical zero-age horizontal branch (ZAHB) 
due to a combination of lower \teff\ and/or higher \logg.
The three stars between 10,000~K and 12,000~K, however, fall {\em below}
the canonical ZAHB when fitted with enriched model atmospheres. This is
plausible as the radiative levitation is supposed to start around 11,000 --
12,000~K (Grundahl et al.\ \cite{grca99}) and the cooler stars therefore
should have metal-poor atmospheres (see also Fig.~\ref{n6752bhb_abu}, where
the coolest analysed star shows no evidence of iron enrichment). This
assumption is also supported by the results of Glaspey et al.\
(\cite{glde85}, NGC~6397; \cite{glmi89}, NGC~6752) and Behr et al.\
(\cite{beco99}, M~13; \cite{beco00b}, M~15). Now the stars below 15,300~K
scatter around the locus defined
by the canonical HB tracks. The stars between 15,500~K and 19,000~K,
however, still show offsets from the canonical locus while for the sdB
stars not much is changed. 
Interestingly, 15,500~K is roughly the temperature\footnote{We determined this
temperature by comparing the $(u-y)_0$ value, at which the stars return to
the ZAHB ($(u-y)_0 \approx +0.4$) to theoretical colours from Kurucz
(\cite{kuru92}) for [M/H] = $+0.5$, which is the metallicity required to
explain the $u$-jump. Assuming \logg\ = 4.0 this comparison
results in \teff\ $\approx$ 15,000~K. }
at which the stars in NGC~6752 return to the ZAHB in
($u-y$, $u$) of Grundahl et al.\ (\cite{grca99}). 
Grundahl et al. caution, however, that their faint 
photometry for NGC 6752 might be affected by poor seeing, and that in the 
Str\"omgren CMD of the better observed cluster, M13, the stars do not return to 
the ZAHB until a temperature of about 20,000 K.

\begin{table}
\caption[Physical parameters, helium abundances, and masses for targets in
NGC~6752 using solar-metallicity model atmospheres] {Physical parameters,
helium abundances, and masses for the target stars in NGC~6752 as derived
using solar metallicity model atmospheres. \label{n6752bhb_parp00}} 
\begin{tabular}{l|lrrr}
\hline
Star & \multicolumn{1}{c}{\teff} & \multicolumn{1}{c}{\logg} &
 \multicolumn{1}{c}{\loghe} & \multicolumn{1}{c}{M}\\
 & \multicolumn{1}{c}{[K]} & \multicolumn{1}{c}{[cm s$^{-2}$]} & &
\multicolumn{1}{c}{[\Msolar]} \\
\hline
\multicolumn{5}{c}{ESO 1.52m telescope observations in 1998}\\
\hline
  652 & 12500$\pm$230 &  3.98$\pm$0.07 & $-$2.19$\pm$0.36 & 0.67\\
 1132 & 16300$\pm$460 &  4.31$\pm$0.07 & $-$2.45$\pm$0.16 & 0.51\\
 1152 & 15000$\pm$290 &  4.21$\pm$0.05 & $-$2.58$\pm$0.17 & 0.52\\
 1157 & 15000$\pm$360 &  4.15$\pm$0.07 & $-$2.89$\pm$0.31 & 0.52\\
 1738 & 15800$\pm$580 &  4.15$\pm$0.10 & $-$2.23$\pm$0.28 & 0.32\\
 2735 & 11400$\pm$170 &  3.96$\pm$0.07 & $-$1.51$\pm$0.28 & 0.98\\
 3253 & 13500$\pm$310 &  3.88$\pm$0.07 & $-$2.57$\pm$0.28 & 0.57\\
 3348 & 12100$\pm$220 &  3.86$\pm$0.07 & $-$2.44$\pm$0.38 & 0.80\\
 3408 & 14100$\pm$330 &  4.24$\pm$0.07 & $-$2.47$\pm$0.36 & 0.66\\
 3410 & 14900$\pm$370 &  4.17$\pm$0.07 & $-$2.25$\pm$0.19 & 0.43\\
 3424 & 16800$\pm$510 &  4.21$\pm$0.09 & $-$2.58$\pm$0.22 & 0.43\\
 3450 & 13000$\pm$210 &  3.92$\pm$0.05 & $-$2.22$\pm$0.24 & 0.49\\
 3461 & 14600$\pm$400 &  4.20$\pm$0.09 & $\le-$3 & 0.73\\
 3655 & 24900$\pm$1250 &  5.14$\pm$0.16 & $-$2.32$\pm$0.24 & 0.65\\
 3736 & 13200$\pm$270 &  3.99$\pm$0.07 & $-$1.98$\pm$0.17 & 0.77\\
 4172 & 12300$\pm$200 &  3.81$\pm$0.05 & $-$2.49$\pm$0.55 & 0.57\\
 4424 & 12900$\pm$210 &  4.07$\pm$0.05 & $-$2.58$\pm$0.40 & 0.78\\
 4551 & 14900$\pm$410 &  3.99$\pm$0.09 & $-$2.26$\pm$0.24 & 0.41\\
 4822 & 13600$\pm$350 &  3.97$\pm$0.09 & $-$2.37$\pm$0.28 & 0.41\\
 4951 & 16300$\pm$520 &  4.38$\pm$0.09 & $-$2.61$\pm$0.22 & 0.57\\
\hline
\multicolumn{5}{c}{ESO NTT observations in 1997}\\ 
\hline
  944 & 11400$\pm$190 &  3.90$\pm$0.09 & $-$1.27$\pm$0.22 & 0.74\\
 1391 & 18500$\pm$570 &  4.45$\pm$0.09 & $-$2.02$\pm$0.10 & 0.36\\
 1780 & 16900$\pm$530 &  4.37$\pm$0.09 & $-$2.28$\pm$0.14 & 0.37\\
 2099 & 18800$\pm$790 &  4.58$\pm$0.10 & $-$2.36$\pm$0.22 & 0.46\\
\hline
\multicolumn{5}{c}{ESO NTT observations in 1998 }\\
\hline
 2697 & 15000$\pm$360 &  4.10$\pm$0.07 & $-$2.39$\pm$0.17 & 0.94\\
 2698 & 14700$\pm$490 &  4.13$\pm$0.10 & $-$2.10$\pm$0.26 & 0.50\\
 2747 & 21600$\pm$700 &  4.80$\pm$0.09 & $-$2.16$\pm$0.10 & 0.55\\
 2932 & 17500$\pm$600 &  4.61$\pm$0.10 & $-$1.54$\pm$0.10 & 0.46\\
 3006 & 29100$\pm$740 &  5.18$\pm$0.09 & $\le-$3 & 0.68\\
 3094 & 10800$\pm$310 &  4.01$\pm$0.14 & $-$2.33$\pm$1.97 & 1.52\\
 3253 & 13300$\pm$370 &  3.81$\pm$0.09 & $-$1.95$\pm$0.31 & 0.50\\
 3699 & 21800$\pm$1050 &  4.60$\pm$0.12 & $-$2.30$\pm$0.10 & 0.32\\
\hline
\multicolumn{5}{c}{ESO NTT observations in 1993}\\
\hline
  491 & 28100$\pm$540 &  5.40$\pm$0.07 & $\le-$3 & 0.37\\
  916 & 29400$\pm$480 &  5.60$\pm$0.07 & $-$1.70$\pm$0.05 & 0.46\\
 1509 & 16400$\pm$510 &  4.07$\pm$0.09 & $-$2.15$\pm$0.16 & 0.24\\
 1628 & 20600$\pm$620 &  4.78$\pm$0.09 & $-$2.52$\pm$0.12 & 0.43\\
 2162 & 33400$\pm$460 &  5.79$\pm$0.07 & $-$1.92$\pm$0.09 & 0.44\\
 2395 & 21000$\pm$750 &  5.06$\pm$0.10 & $-$1.78$\pm$0.09 & 0.53\\
 3915 & 30700$\pm$620 &  5.54$\pm$0.09 & $\le-$3 & 0.56\\
 3975 & 20400$\pm$520 &  4.92$\pm$0.07 & $-$2.02$\pm$0.12 & 0.62\\
 4009 & 30100$\pm$1120 &  5.60$\pm$0.14 & $\le-$3 & 0.52\\
 4548 & 20700$\pm$1490 &  5.06$\pm$0.19 & $-$2.00$\pm$0.17 & 0.62\\
\hline
\end{tabular}
\end{table}

\begin{table}
\caption[Iron and magnesium abundances for the HBB stars in NGC~6752]
{Helium, iron, and magnesium abundances of the HBB stars observed with the
ESO 1.52m telescope (except B~3655, which has a too noisy spectrum). [Fe/H]
and [Mg/H] are derived using solar abundances of $\log \epsilon_{\rm Fe,
\odot} = 7.46$ and $\log \epsilon_{\rm Mg, \odot} = 7.53$. The physical
parameters and the helium abundances
are taken from Table~\ref{n6752bhb_parp00}. \label{irontab}} 
\begin{tabular}{lrrrrr}
\hline
Star & \teff & \logg & \loghe & [Fe/H] & [Mg/H]\\
     & [K]   & [cm s$^{-2}$] & & & \\
\hline
  652 & 12500 &  3.98 & $-$2.19 & $+$0.1 & $-$0.9\\ 
 1132 & 16300 &  4.31 & $-$2.45 & $+$0.5 & $-$1.0\\ 
 1152 & 15000 &  4.21 & $-$2.58 & $+$0.2 & $-$1.4\\ 
 1157 & 15000 &  4.15 & $-$2.89 & $+$0.2 & $-$1.2\\ 
 1738 & 15800 &  4.15 & $-$2.23 & $+$0.2 & $-$0.9\\ 
 2735 & 11400 &  3.96 & $-$1.51 & $<-$1.6& $-$1.5\\ 
 3253 & 13500 &  3.88 & $-$2.57 & $+$0.8 & $-$1.3\\ 
 3348 & 12100 &  3.86 & $-$2.44 & $-$0.2 & $-$1.1\\ 
 3408 & 14100 &  4.24 & $-$2.47 & $+$0.2 & $-$1.3\\ 
 3410 & 14900 &  4.17 & $-$2.25 & $+$0.6 & $-$1.5\\ 
 3424 & 16800 &  4.21 & $-$2.58 & $-$0.1 & $-$1.2\\ 
 3450 & 13000 &  3.92 & $-$2.22 & $-$0.4 & $-$1.2\\ 
 3461 & 14600 &  4.20 & $\le-$3 & $+$0.3 & $-$0.7\\ 
 3736 & 13200 &  3.99 & $-$1.98 & $+$0.3 & $-$0.9\\ 
 4172 & 12300 &  3.81 & $-$2.49 & $-$0.2 & $-$1.7\\ 
 4424 & 12900 &  4.07 & $-$2.58 & $+$0.6 & $-$1.5\\ 
 4551 & 14900 &  3.99 & $-$2.26 & $-$0.9 & $-$0.6\\ 
 4822 & 13600 &  3.97 & $-$2.37 & $-$0.2 & $-$0.9\\ 
 4951 & 16300 &  4.38 & $-$2.61 & $+$0.2 & $-$1.0\\ 
\hline                                    
\end{tabular}
\end{table}

We next repeated the Balmer line profile fits by increasing the metal
abundance of the model atmospheres to [M/H]=$+$0.5 (see
Fig.~\ref{n6752bhb_irontg}, bottom panel, and Table~\ref{n6752bhb_parp05}),
which did not significantly change the resulting values for \teff\ and
\logg . In particular, note that especially
the ``deviant'' stars (now between 15,300~K and 19,000~K)
remain offset from the canonical ZAHB. 

\begin{table}
\caption{Physical parameters, helium abundances, and masses for
the target stars in NGC~6752 as derived using metal-rich model
atmospheres. \label{n6752bhb_parp05}}
\begin{tabular}{l|lrrr}
\hline
Star & \multicolumn{1}{c}{\teff} & \multicolumn{1}{c}{\logg} &
 \multicolumn{1}{c}{\loghe} & \multicolumn{1}{c}{M}\\
 & \multicolumn{1}{c}{[K]} & \multicolumn{1}{c}{[cm s$^{-2}$]} & &
\multicolumn{1}{c}{[\Msolar]} \\
\hline
\multicolumn{5}{c}{ESO 1.52m telescope observations in 1998}\\
\hline
  652 & 12700$\pm$220 & 4.05$\pm$0.07 & $-$2.40$\pm$0.36 & 0.73\\
 1132 & 16300$\pm$430 & 4.36$\pm$0.07 & $-$2.55$\pm$0.14 & 0.54\\
 1152 & 15100$\pm$290 & 4.26$\pm$0.05 & $-$2.71$\pm$0.16 & 0.55\\
 1157 & 15100$\pm$370 & 4.20$\pm$0.07 & $-$2.98$\pm$0.26 & 0.55\\
 1738 & 15900$\pm$540 & 4.21$\pm$0.10 & $-$2.37$\pm$0.26 & 0.35\\
 2735 & 11600$\pm$180 & 4.08$\pm$0.07 & $-$1.74$\pm$0.24 & 1.20\\
 3253 & 13700$\pm$300 & 3.94$\pm$0.07 & $-$2.76$\pm$0.24 & 0.61\\
 3348 & 12300$\pm$210 & 3.94$\pm$0.07 & $-$2.65$\pm$0.36 & 0.89\\
 3408 & 14200$\pm$310 & 4.30$\pm$0.07 & $-$2.64$\pm$0.35 & 0.71\\
 3410 & 15000$\pm$350 & 4.23$\pm$0.07 & $-$2.40$\pm$0.19 & 0.47\\
 3424 & 16700$\pm$490 & 4.24$\pm$0.09 & $-$2.66$\pm$0.21 & 0.43\\
 3450 & 13200$\pm$200 & 3.99$\pm$0.05 & $-$2.45$\pm$0.26 & 0.53\\
 3461 & 14700$\pm$380 & 4.26$\pm$0.09 & $-$3.37$\pm$0.33 & 0.79\\
 3655 & 25000$\pm$170 & 5.16$\pm$0.16 & $-$2.31$\pm$0.24 & 0.64\\
 3736 & 13400$\pm$260 & 4.07$\pm$0.07 & $-$2.17$\pm$0.17 & 0.86\\
 4172 & 12500$\pm$200 & 3.88$\pm$0.05 & $-$2.70$\pm$0.50 & 0.62\\
 4424 & 13000$\pm$210 & 4.13$\pm$0.05 & $-$2.79$\pm$0.35 & 0.84\\
 4551 & 15000$\pm$400 & 4.05$\pm$0.09 & $-$2.43$\pm$0.24 & 0.44\\
 4822 & 13800$\pm$340 & 4.04$\pm$0.09 & $-$2.59$\pm$0.28 & 0.44\\
 4951 & 16300$\pm$480 & 4.42$\pm$0.09 & $-$2.72$\pm$0.21 & 0.60\\
\hline
\multicolumn{5}{c}{ESO NTT observations in 1997} \\
\hline
  944 & 11600$\pm$180 & 4.00$\pm$0.07 & $-$1.52$\pm$0.19 & 0.87\\
 1391 & 18400$\pm$580 & 4.48$\pm$0.09 & $-$2.08$\pm$0.10 & 0.37\\
 1780 & 16900$\pm$500 & 4.41$\pm$0.09 & $-$2.37$\pm$0.12 & 0.39\\
 2099 & 18700$\pm$810 & 4.60$\pm$0.10 & $-$2.41$\pm$0.22 & 0.46\\
\hline
\multicolumn{5}{c}{ESO NTT observations in 1998}\\
\hline
 2697 & 15000$\pm$330 & 4.14$\pm$0.07 & $-$2.52$\pm$0.17 & 0.98\\
 2698 & 14800$\pm$490 & 4.20$\pm$0.10 & $-$2.25$\pm$0.28 & 0.55\\
 2747 & 21500$\pm$830 & 4.81$\pm$0.09 & $-$2.17$\pm$0.10 & 0.54\\
 2932 & 17300$\pm$570 & 4.65$\pm$0.10 & $-$1.61$\pm$0.10 & 0.49\\
 3006 & 29300$\pm$590 & 5.19$\pm$0.07 & $-$3.01$\pm$0.31 & 0.65\\
 3094 & 11100$\pm$290 & 4.14$\pm$0.10 & $-$2.54$\pm$1.97 & 1.87\\
 3253 & 13500$\pm$350 & 3.87$\pm$0.09 & $-$2.12$\pm$0.31 & 0.54\\ 
 3699 & 21800$\pm$030 & 4.61$\pm$0.12 & $-$2.31$\pm$0.10 & 0.31\\
\hline
\multicolumn{5}{c}{ESO NTT observations in 1993}\\
\hline
  491 & 28000$\pm$520 & 5.40$\pm$0.07 & $-$3.40$\pm$0.14 & 0.35\\
  916 & 29300$\pm$460 & 5.59$\pm$0.05 & $-$1.70$\pm$0.05 & 0.43\\
 1509 & 16400$\pm$500 & 4.11$\pm$0.09 & $-$2.25$\pm$0.16 & 0.26\\
 1628 & 20700$\pm$720 & 4.81$\pm$0.09 & $-$2.55$\pm$0.12 & 0.43\\
 2162 & 33400$\pm$500 & 5.78$\pm$0.07 & $-$1.91$\pm$0.09 & 0.41\\
 2395 & 20900$\pm$810 & 5.07$\pm$0.09 & $-$1.80$\pm$0.09 & 0.52\\
 3915 & 30600$\pm$580 & 5.54$\pm$0.07 & $-$3.19$\pm$0.19 & 0.53\\
 3975 & 20300$\pm$580 & 4.94$\pm$0.07 & $-$2.06$\pm$0.12 & 0.62\\
 4009 & 30100$\pm$040 & 5.62$\pm$0.12 & $-$3.14$\pm$0.19 & 0.51\\
 4548 & 20400$\pm$590 & 5.06$\pm$0.19 & $-$2.03$\pm$0.17 & 0.60\\
\hline
\end{tabular}
\end{table}

\subsection{Helium mixing\label{ssec-hemix}}

As outlined in Sect.~\ref{n6752bhb_sec_intro}, 
helium mixing during the
RGB phase may also be able to explain the low gravities of
the HBB stars.  Under this scenario the mixing currents
within the radiative zone below the base of the convective
envelope of a red giant star are assumed to penetrate into
the top of the hydrogen shell where helium is being produced
by the hydrogen burning reactions.  Ordinarily one would
expect the gradient in the mean molecular weight $\mu$ to
prevent any penetration of the mixing currents into the
shell. If, however, the timescale for mixing were shorter
than the timescale for nuclear burning, then the helium
being produced at the top of the shell might be mixed
outward into the envelope before a $\mu$ gradient is
established.  Under these circumstances a $\mu$ gradient would
not inhibit deep mixing simply because such a gradient would
not exist within the mixed region.

Since deep mixing is presumably driven by rotation, one
would expect a more rapidly rotating red giant to show a
larger increase in the envelope helium abundance. This, in
turn, would lead to a brighter RGB tip luminosity and hence
to greater mass loss. The progeny of the more rapidly
rotating giants should therefore lie at higher effective
temperatures along the HB than the progeny of the more
slowly rotating giants. This predicted increase in the
stellar rotational velocity with effective temperature along
the HB has not, however, been confirmed by the recent
observations of M13 by Behr et al. (\cite{bedj00a}).  These
observations show that HB stars in M13 hotter than 11,000 K
are, in fact, rotating slowly with $v \sin i <$ 10 km s$^{-1}$ in
contrast to the cooler HB stars where rotational velocities
as high as 40 km s$^{-1}$ are found (see also Peterson
et al.\ \cite{pero95}).

There are a couple of possible explanations for this
apparent discrepancy.  One possibility is that the greater
mass loss suffered by the HBB stars might carry away so much
angular momentum that the surface layers are spun down even
though the core is still rotating rapidly. Alternatively
Sills \& Pinsonneault (\cite{sipi00}) have suggested that the
observed gravitational settling of helium in HBB stars might
set up a $\mu$ gradient in the outer layers which inhibits the
transfer of angular momentum from the rapidly rotating
interior to the surface. Thus the surface rotational
velocities may not necessarily be indicative of the interior
rotation.

In order to explore the consequences of helium
mixing for the HBB stars quantitatively, we evolved a set of 13 sequences
up the RGB to the helium flash for varying amounts of helium mixing using
the approach of Sweigart (\cite{swei97a}, \cite{swei97b}).  As
in the case of the canonical models discussed previously,
all of these mixed sequences had an
initial helium abundance Y of 0.23 and a scaled-solar metallicity [M/H] of
$-$1.54.  The main-sequence mass was taken to be 0.805~\Msolar,
corresponding to an age at the tip of the RGB of 15 Gyr. The mixing depth,
as defined by the parameter $\Delta X_{mix}$ of Sweigart (\cite{swei97a},
\cite{swei97b}),
ranged from 0.0 (canonical, unmixed case) to 0.24 in increments of 0.02. 
Mass loss via the Reimers formulation was included in the calculations with
the mass-loss parameter $\eta_R$ set equal to 0.40.  This value
for $\eta_R$ was chosen so that a canonical, unmixed model
would lie near the red end of the observed blue 
HB in NGC~6752. Both the mixing and
mass loss were turned off once the models reached the
core He flash at the tip of the RGB, and
the subsequent evolution was then followed through the helium flash to the
end of the HB phase using standard techniques. 

We did not investigate the changes in the surface abundances
of CNO, Na and Al caused by the helium mixing, since such a
study was beyond the scope of the present paper.  Rather,
our objective was to determine how the mixing affected those
quantities which impact on the HB evolution, i.e., envelope
helium abundance and mass.  We do note that the mixing in
the more deeply mixed RGB models would have penetrated into
regions of substantial Na and Al production according to the
calculations of Cavallo et al. (\cite{casw96}, \cite{casw98}).  However, the
resulting changes in the surface Na and Al abundances will
depend on the assumed initial Ne and Mg isotopic abundances
and on the adopted nuclear reaction rates, which in some
cases are quite uncertain.

The locus of the above helium-mixed sequences in the \logg\ - \logt\ plane is
indicated by the dashed lines in the top panel of Fig.~4.  The red end of
the mixed ZAHB in this panel, located at \logt\  = 3.93, is set by the
canonical, unmixed sequence for the present set of model parameters.  Since
mixing increases the RGB mass loss, a mixed HB model will
have a higher effective temperature than the corresponding canonical
model. At the same time mixing
increases the envelope helium abundance in the HB model, which, in turn,
increases both the hydrogen-burning and surface luminosities.  The net
effect is to shift the mixed locus in Fig.~\ref{n6752bhb_irontg}  towards
lower gravities with increasing \teff\ compared to the canonical locus,
until a maximum offset is reached for 15,500~K $<$ \teff\ $<$ 19,000~K.  At
higher temperatures the mixed locus shifts back towards the canonical
locus, as the contribution of the hydrogen shell to the surface luminosity
declines due to the decreasing envelope mass.  The predicted locus along
the extreme HB (EHB) 
does not depend strongly on the extent of the mixing, since the
luminosities and gravities of the EHB stars are primarily determined by the
mass of the helium core, which is nearly the same for the mixed and
canonical models.  Overall the variation of \logg\ with \teff\ along the
mixed locus in the top panel of Fig.~\ref{n6752bhb_irontg}  mimics the
observed variation. 

The results presented in Sect.~\ref{n6752bhb_ssec_iron}  demonstrate that
radiative levitation of heavy elements can account for a considerable
fraction of the gravity offset along the HBB, especially for temperatures
cooler than 15,100~K.  Consequently the amount of helium mixing required to
explain the remaining offset between 15,300~K and 19,000~K is much less
than the amount required to explain the offsets found without 
accounting for radiative
levitation (top panel of Fig.~\ref{n6752bhb_irontg}). In order to
compare the gravities predicted by the helium-mixing scenario with those
derived from the metal-enhanced atmospheres, we computed a second set of
mixed sequences using the same approach as above but with a larger value of
the mass-loss parameter $\eta_R$, i.e., $\eta_R$ = 0.45.  
The red end of the mixed ZAHB for these sequences is located at
\logt\ = 4.01 and is therefore hotter than the red end
of the mixed ZAHB for the sequences with $\eta_R$ = 0.40. The HB stars
cooler than this temperature in NGC~6752 would then be identified
with unmixed stars which lost less mass along the RGB.

By increasing the
mass loss efficiency we reduce the amount of mixing needed to populate the
temperature range 15,300 K $<$ \teff\ $<$ 19,000 K and therefore the size
of the resulting gravity offset.  The locus of the mixed sequences with
$\eta_R$ = 0.45 is indicated by the dashed lines in the central panel of
Fig.~\ref{n6752bhb_irontg}. The gravity offsets along this mixed locus seem
to provide a reasonable fit to the gravities given by the model atmospheres
with solar metallicity.

Finally we computed a third set of mixed sequences with the mass-loss
parameter increased further to $\eta_R$ = 0.50 for comparison with the
gravities obtained from the atmospheres with super-solar metallicity in the
bottom panel of Fig.~\ref{n6752bhb_irontg}.  As expected, these mixed
sequences show a smaller gravity offset in the temperature range 15,500~K
$<$ \teff\ $<$ 19,000~K. Moreover, 
the red end of the mixed ZAHB shifts blueward to \logt\ = 4.08.

\section{Masses}
We calculated masses for the programme stars in NGC~6752
from their values of \teff\ and \logg\ using the equation: 

\noindent
$\log{\frac{\rm M}{\rm M_\odot}} = const.+\log g+0.4\cdot((m-M)_V-V+V_{th})$

\noindent
where $V_{th}$ denotes the theoretical brightness at the stellar surface as
given by Kurucz (\cite{kuru92}). We decided to use the photometry of
Buonanno et al.\ (\cite{buca86}) to derive masses. As can be seen from
Table~\ref{n6752bhb_targ} the photometry of Thompson et al.\ (\cite{thka99})
yields in general fainter visual magnitudes than the photometry of Buonanno
et al.\ The effect on the masses, however, is small: On average, the masses
derived from the Thompson et al. photometry are 5\% lower than those derived
from the Buonanno et al. photometry. We adopted $(m-M)_0 = 13.17$ and \ebv\
= 0.04 for the distance modulus and reddening. These are mean values
derived from the determinations of Renzini et al.\ (\cite{rebr96}), Reid
(\cite{reid97}, \cite{reid98}), and Gratton et al.\ (\cite{grfu97}). The
errors in \logM\ are estimated to be about the same as obtained for the
older NGC~6752 data described by Moehler et al.\ (\cite{mohe97b}): 0.15 dex
for stars above the gap, 0.17 dex for stars within the gap region and 0.22
dex for stars below the gap. 

\begin{figure}
\vspace{11.5cm}
\includegraphics{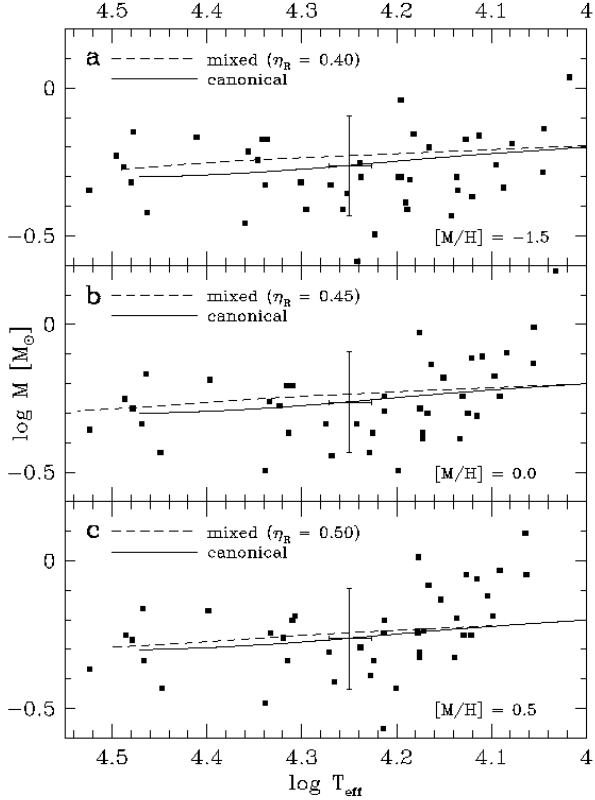}
\caption[\teff, $\log M/M_\odot$ for stars in NGC~6752]
{{\bf a-c} Temperatures and masses (derived from Buonanno et al.'s photometry)
of the programme stars in NGC~6752.
{\bf a} determined using model atmospheres with cluster 
metallicity ([M/H] = $-$1.5),
{\bf b} adopting a solar metallicity ([M/H] = 0) for the model
atmospheres
{\bf c} adopting a super--solar metallicity ([M/H] = $+$0.5)
for the model atmospheres.
For more details see Sect.~\ref{n6752bhb_ssec_iron}. 
The dashed resp. solid lines mark the ZAHB masses
for a metallicity [M/H]~=~$-$1.54, as computed with and
without mixing, respectively. 
(see Sect.~\ref{ssec-hemix} and Fig.~\ref{n6752bhb_irontg} for details).
\label{n6752bhb_irontlm}} 
\end{figure}

\begin{table*}
\caption[]{Mean mass ratios between spectroscopically derived masses and
predicted zero-age HB masses at the same effective temperatures. 
B~2697, B~3006 and stars cooler than 11,500~K are excluded from this 
comparison. $\eta_{\rm R}$ gives
the Reimers' mass loss parameter for the respective ZAHB. We derived the
masses using the photometry of Buonanno et al.\ (\cite{buca86}). The cited
errors are standard deviations. \label{mass_zahb}} 
\begin{tabular}{lll|c|l}
\hline
 cool HBB stars & hot HBB stars & sdB stars & [M/H] & track \\
\hline
   $0.89^{+0.20}_{-0.16}$ (16) & $0.76^{+0.21}_{-0.17}$ (9) 
 & $1.04^{+0.25}_{-0.20}$ (12) & $-$1.5 
 & canonical HB, variable $\eta_{\rm R}$ \\
   $0.84^{+0.19}_{-0.16}$ (16) & $0.70^{+0.19}_{-0.15}$ (9)
 & $0.95^{+0.23}_{-0.18}$ (12) & $-$1.5 
 & mixed HB, $\eta_{\rm R}$ = 0.40\\
\hline
   $0.97^{+0.24}_{-0.19}$ (16)  & $0.73^{+0.22}_{-0.17}$ (9)   
 & $0.97^{+0.23}_{-0.19}$ (12)  & $+$0.0 
 & canonical HB, variable $\eta_{\rm R}$ \\
   $0.94^{+0.23}_{-0.19}$ (16) & $0.69^{+0.20}_{-0.16}$ (9)
 & $0.91^{+0.22}_{-0.18}$ (12) & $+$0.0 
 & mixed HB, $\eta_{\rm R}$ = 0.45\\
\hline
  $1.06^{+0.27}_{-0.21}$ (16) & $0.76^{+0.22}_{-0.17}$ (9) 
 & $0.95^{+0.24}_{-0.19}$ (12) & $+$0.5 
 & canonical HB, variable $\eta_{\rm R}$ \\
  $1.04^{+0.27}_{-0.21}$ (16) & $0.73^{+0.21}_{-0.16}$ (9)
 & $0.90^{+0.22}_{-0.18}$ (12) & $+$0.5 
 & mixed HB, $\eta_{\rm R}$ = 0.50\\
\hline
\end{tabular}
\end{table*}

The masses derived from the analysis using metal-poor model atmospheres are
plotted in Fig.~\ref{n6752bhb_irontlm} (top panel). The sdB stars hotter
than 20,000~K (\logt\ = 4.3) scatter around the canonical ZAHB, 
whereas the stars
below 16,000~K (\logt\ = 4.2) lie mainly below the canonical
ZAHB. Even stronger
deviations towards low masses are found between 16,000~K and 20,000~K.
Comparing the masses to those predicted by the mixed ZAHB ($\eta_{\rm R} =$
0.40) we obtain similar results. To quantify the offsets we compare the
masses of the stars to those they would have on the theoretical ZAHB at the
same \teff\ (see Table~\ref{mass_zahb}). We divide the stars into
three groups for the further discussion (excluding stars below 11,500~K,
for which diffusion should play no r\^ole, as well as B~2697 and B~3006):
cool HBB stars (\teff\ $<$ 16,000~K, 16 stars), hot HBB stars (16,000~K
$\le$ \teff\ $<$ 20,000~K, 9 stars), and sdB stars (\teff\ $\ge$ 20,000~K,
12 stars). The effective temperatures here are those derived from
metal-poor model atmospheres.  

The results of the analyses using solar-metallicity model atmospheres are
plotted in the central panel of Fig.~\ref{n6752bhb_irontlm} (see
Table~\ref{mass_zahb}). The effect on the masses is similar to that on the
temperatures/gravities (Fig.~\ref{n6752bhb_irontg}, central panel) -- below
15,300~K (cool HBB stars) and above 20,000~K (sdB stars) the masses
basically scatter around the canonical ZAHB, but the hot HBB stars between
these two groups still show too low masses. 
Comparing the masses to those predicted by the
mixed ZAHB ($\eta_{\rm R}$ = 0.45) gives similar results. As the stars
become cooler when analysed with more metal-rich atmospheres the
temperature boundaries were shifted to include the same stars as for the
comparison made above. 

Even the use of metal-rich model atmospheres for the analyses does not
change much (see Table~\ref{mass_zahb} and the bottom panel of
Fig.~\ref{n6752bhb_irontlm}). Obviously the hot HBB stars in the
intermediate temperature range still show low masses, despite the use of
metal-rich model atmospheres. Thus the problem of the stars in this
temperature range(15,300~K$<$ \teff\ $<$19,000~K)  cannot be completely
solved by the scaled-solar metal-rich atmospheres used here. 

\section{Discussion}

We find that the atmospheres
of HBB stars in NGC~6752 with \teff\ $>$ 11,500~K are enriched in iron 
([Fe/H] $\approx+$0.1) whereas their magnesium abundances are the same as 
found in cluster giants. Our results are consistent with those of Behr et 
al. (\cite{beco99}, \cite{beco00b}) for HBB stars in M~13 and M~15. Using
model atmospheres that try to take into account this enrichment in iron
(and presumably other heavy elements) reconciles the atmospheric parameters
of stars with 11,500~K $\le$ \teff\ $<$ 15,100~K with canonical
expectations as suggested by Grundahl et al.\ (\cite{grca99}). Also the
masses derived from these analyses are in good agreement with canonical
predictions within this temperature range. However, we found that even with
model atmospheres as metal rich as [M/H] = $+$0.5 the atmospheric
parameters of the hot HBB stars (15,300~K $<$ \teff\ $<$ 19,000~K) in
NGC~6752 cannot be reconciled with the canonical ZAHB. Both the gravities
and masses of these hot HBB stars remain too low. In addition, the masses
for the stars below 15,100~K are slightly too high for the super-solar 
metallicity (the EHB stars are
hardly affected at all by changes in the metallicity of the model
atmospheres). 
             
Michaud et al.\ (\cite{miva83}) noted that diffusion will not necessarily
enhance all heavy elements by the same amount and that the effects of
diffusion vary with effective temperature. Elements that were originally
very rare may be enhanced even more strongly than iron (see also Behr et
al.\ \cite{beco99}, where P and Cr are enhanced to [M/H] $\ge +1$). The
question of whether diffusion is {\em the} (one and only) solution to the
``low gravity'' problem cannot be answered without detailed abundance
analyses to determine the actual abundances and the use of model
atmospheres that allow the use of non-scaled solar abundances (like
ATLAS12). We can, however, state that those model atmospheres, which
reproduce the $u$-jump discussed by Grundahl et al.\ (\cite{grca99}) cannot
completely reconcile the atmospheric parameters of hot HB stars with
canonical theory. Model atmospheres with abundance distributions that may
solve the discrepancy between theoretically predicted and observed
atmospheric parameters of hot HB stars may then, in turn, not reproduce the
Str\"omgren $u$-jump. It is intriguing that the temperature, at which the
stars in $u, u-y$ seem to return to the ZAHB, is roughly the same at which
they start to deviate again from the canonical ZAHB in \logg, \teff\ when
analysed with metal-rich atmospheres. 

The stars between 15,300~K and 19,000~K (when analysed with metal-rich
atmospheres) are currently best fit by a moderately mixed ZAHB. However,
the fact that their masses are too low cautions against identifying He
mixing as the only cause for these low gravities - because in this case the
luminosities of the stars would be increased and canonical masses would
result. 


\acknowledgements

We want to thank the staff of the ESO La Silla observatory for their support 
during our observations. We would also like to thank T. Lanz for very
helpful discussions and the referees J. Cohen and B. Behr for useful 
remarks. S.M. acknowledges financial support from the DARA
under grant 50~OR~96029-ZA. A.V.S. acknowledges 
financial support from NASA Astrophysics Theory Program proposal 
NRA-99-01-ATP-039.

{}

\begin{thebibliography}{}
\bibitem[1999]{beco99}
Behr B.B., Cohen J.G., McCarthy J.K., Djorgovski S.G., 1999, ApJ 517, L135
\bibitem[2000a]{bedj00a}
Behr B.B., Djorgovski S.G., Cohen J.G., McCarthy J.K., C\^ol\'e P.,
Piotto G., Zoccali M., 2000a, ApJ 528, 849
\bibitem[2000b]{beco00b}
Behr B.B., Cohen J.G., McCarthy J.K., 2000b, ApJ 531, L37
\bibitem[1986]{buca86}
Buonanno R., Caloi V., Castellani V., et al., 1986, A\&AS 66, 79
\bibitem[1992]{besa92}
 Bergeron P., Saffer R.A., Liebert J., 1992, ApJ 394, 228
\bibitem[1999]{calo99}
 Caloi, V., 1999, A\&A 343, 904 
\bibitem[1999]{cagr99}
Carretta E., Gratton R., Sneden C., Bragaglia A., 2000, A\&A 354, 169
\bibitem[1996]{casw96}
Cavallo R.M., Sweigart A.V., Bell R.A., 1996, ApJ 464, L79
\bibitem[1998]{casw98}
Cavallo R.M., Sweigart A.V., Bell R.A., 1998, ApJ 492, 575
\bibitem[1995]{char95}
Charbonnel C., 1995, ApJ 453, L41
\bibitem[1999]{fuai99}
Fujimoto M.Y., Aikawa M., Kato K., 1999, ApJ 519, 733
\bibitem[1985]{glde85}
Glaspey J.W., Demers S., Moffat A.F.J., Shara M., 1985, ApJ 289, 326
\bibitem[1989]{glmi89}
Glaspey J.W., Michaud G., Moffat A.F.J., Demers S., 1989, ApJ 339, 926
\bibitem[1997]{grfu97}
Gratton R.G., Fusi Pecci F., Carretta E., et al., 1997, ApJ 491, 749
\bibitem[1999]{grca99}
 Grundahl F., Catelan M., Landsman W.B., Stetson P.B., Andersen M., 1999, 
   ApJ 524, 242
\bibitem[1992]{hamu92}
 Hamuy M., Walker A.R., Suntzeff N.B., et al., 1992, PASP 104, 533
\bibitem[1998]{hasn98}
Hanson R.B., Sneden C., Kraft R.P., Fulbright J., 1998, AJ 116, 1286
\bibitem[1997]{hemo97}
Heber U., Moehler S., Reid I.N., 1997, Masses and gravities of blue 
horizontal branch (BHB) stars revisited, In: Battrick B. (ed)
HIPPARCOS Venice '97, ESA SP-402, p. 461
\bibitem[1986]{horn86}
 Horne K., 1986, PASP 98, 609
\bibitem[1994]{kraf94}
 Kraft R.P., 1994, PASP 106, 553
\bibitem[1997]{krsn97}
Kraft R.P., Sneden C., Smith G.H., et al., 1997, AJ 113,279
\bibitem[1991]{kuru91}
 Kurucz R.L., 1991, private communication
\bibitem[1992]{kuru92}
Kurucz R.L., 1992, in {\it The Stellar Populations of Galaxies},
    eds. B. Barbuy \& A. Renzini, IAU Symp. 149 (Kluwer:Dordrecht), 225
\bibitem[1995]{laho95}
Langer G.E., Hoffman R.D., 1995, PASP 107, 1177
\bibitem[1997]{lema97}
Leone F., Manfr\`e M., 1997, A\&A 320, 257
\bibitem[1983]{miva83}
Michaud G., Vauclair G., Vauclair S., 1983, ApJ 267, 256
\bibitem[1998]{misa98}
Mitchell K.J., Saffer R.A., Howell S.B., Brown T.M., 1998, MNRAS 295, 225
\bibitem[1999]{moeh99}
 Moehler S., 1999, RvMA 12, p. 281
\bibitem[1997a]{mohe97a}
Moehler S., Heber U., Durrell P., 1997a, A\&A 317, L83
\bibitem[1997b]{mohe97b}
 Moehler S., Heber U., Rupprecht G., 1997b, A\&A 319, 109 
\bibitem[1998]{mola98}
Moehler S., Landsman W., Napiwotzki R., 1998, A\&A 335, 510
\bibitem[1999a]{mosw99a}
Moehler S., Sweigart A.V., Landsman W.B., Heber U., Catelan M., 1999a, A\&A 
 346, L1
\bibitem[1999b]{mosw99b}
Moehler S., Sweigart A.V., Catelan M., 1999b, A\&A 351, 519
\bibitem[2000]{mola99}
Moehler S., Landsman W., Dorman B., 2000, in prep.
\bibitem[1995a]{noda95a}
Norris J.E., Da Costa G.S., 1995a, ApJ 441, L81
\bibitem[1995b]{noda95b}
Norris J.E., Da Costa G.S., 1995b, ApJ 447, 680
\bibitem[1995]{pero95}
Peterson R.C., Rood R.T., Crocker D.A., 1995, ApJ 453, 214
\bibitem[1997]{reid97}
Reid I.N., 1997, AJ 114, 161
\bibitem[1998]{reid98}
Reid I.N., 1998, AJ 115, 204
\bibitem[1999]{reid99}
Reid I.N., 1999, ARAA 37, 191
\bibitem[1996]{rebr96}
Renzini A., Bragaglia A., Ferraro F.R., et al., 1996, ApJ 465, L23
\bibitem[1994]{sabe94}
 Saffer R.A., Bergeron P., Koester D., Liebert J., 1994, ApJ 432, 351
\bibitem[1997]{sake97}
Saffer R.A., Keenan F.P., Hambly N.C., Dufton P.L., Liebert J., 1997, ApJ 491,
    172
\bibitem[2000]{sipi00}
Sills A., Pinsonneault M.H., 2000, ApJ submitted (astro-ph/9911024)
\bibitem[1997a]{swei97a}
Sweigart A.V., 1997a, ApJ 474, L23
\bibitem[1997b]{swei97b}
 Sweigart A.V., 1997b, Helium Mixing in Globular Cluster Stars. In:
 Philip A.G.D., Liebert J., Saffer R.A. (eds.), The Third Conference
 on Faint Blue Stars. Cambridge University Press, Cambridge, p.~3
\bibitem[1979]{swme79}
Sweigart A.V., Mengel J.G., 1979, ApJ 229, 624
\bibitem[1999]{thka99}
Thompson I.B., Kaluzny J., Pych W., Krzeminski W., 1999, AJ 118, 462
\bibitem[1977]{tueg77}
 T\"ug H., 1977, ESOMe 11, 7
\bibitem[1988]{voru88}
Von Rudloff I.R., VandenBerg D.A., Hartwick F.D.A., 1988, ApJ 324, 840
\bibitem[1992]{zahn92}
Zahn J.P., 1992, A\&A 265, 115
\end{thebibliography}
\end{document}